\documentclass[aps, preprint, amsmath, amssymb]{revtex4-2}

\usepackage{graphicx}
\usepackage{bm}
\usepackage{hyperref}

\begin{document}

\title{Transition to super-diffusive transport in turbulent plasmas}

\author{Matteo Stanzani}
 \altaffiliation[Also at ]{University of Bologna, Dept. of Industrial Engineering - Montecuccolino nuclear engineering laboratory - 40136 Bologna, Italy}
\email{matteo.stanzani3@unibo.it}
\author{Filippo Arlotti}
\altaffiliation[Also at ]{University of Bologna, Dept. of Industrial Engineering - Montecuccolino nuclear engineering laboratory - 40136 Bologna, Italy}
 \email{filippo.arlotti2@studio.unibo.it}
\affiliation{CEA, IRFM, F-13108 Saint-Paul-Lez-Durance, France}

\author{Guido Ciraolo}
\email{guido.ciraolo@cea.fr}
\affiliation{CEA, IRFM, F-13108 Saint-Paul-Lez-Durance, France}

\author{Xavier Garbet}
\email{xavier.garbet@cea.fr}
\affiliation{CEA, IRFM, F-13108 Saint-Paul-Lez-Durance, France}

\author{Cristel Chandre}
\email{cristel.chandre@cnrs.fr}
\affiliation{CNRS, Aix Marseille Univ, I2M, 13009 Marseille, France}

\date{\today}

\begin{abstract}
We investigate the motion of charged particles in a turbulent electrostatic potential using guiding-center theory. By increasing the Larmor radius, the dynamics exhibit close-to-ballistic transport properties. The transition from diffusive to ballistic transport is analyzed using nonlinear dynamics. It is found that twistless invariant tori in the guiding-center dynamics are responsible for this transition, drastically affecting transport properties of charged particles.  
\end{abstract}

\maketitle

Modeling and characterizing transport in magnetically confined plasmas, such as encountered in tokamaks, is a long-standing issue in plasma physics and a prerequisite to the control of turbulence for better confinement properties of the plasma. Several levels of description of charged particle transport are being actively pursued from the more computationally intensive, such as kinetic or gyrokinetic modeling, to the more theoretically palatable theories such as classical or neoclassical theories. The nature of the transport of particles is at the core of these latter theories, and strongly depends on the type of charged particles. For instance, it is expected that the nature of transport for alpha particles is much different than the one for thermal ions due to a large Larmor radius, washing out the fine-scale structures of the electrostatic potential~\cite{Manfredi1996}.

The main objective of this article is to characterize the transport properties in a rather simplified setting which captures some of the main features present in electrostatic turbulence. We use this simplified setting to uncover the phase-space structures organizing the dynamics and responsible for transport properties.  

In this article, we consider a constant and uniform magnetic field to focus on the transport properties caused by electrostatic drift waves. The motion of a charged particle of mass $m$ and charge $q$ in a strong magnetic field ${\bf B}=B\hat{\bf z}$ and a turbulent electrostatic potential $\Phi({\bf x},t)$ is given by
\begin{equation}
\label{eqn:fo}
m\frac{{\rm d} {\bf v}}{{\rm d} t} = q\left( -\nabla \Phi ({\bf x},t) + {\bf v}\times {\bf B}\right),
\end{equation}
where ${\bf x}=(x,y,z)$ and ${\bf v}=(v_x,v_y,v_z)$ are the position and the velocity of the charged particle. We decouple the dynamics along the magnetic field lines (i.e., along the $z$ direction) and perpendicular to the magnetic field lines (i.e., in the $(x,y)$ plane)  by considering that the electrostatic potential $\Phi$ does not depend on the longitudinal coordinate $z$. In the transverse plane, the motion is composed of a fast gyration with Larmor frequency $\Omega =q B/m$ (its sign indicating the rotational direction) and a slower drift motion across magnetic field lines. 
The main question we address is to characterize the slow drift motion as a function of the main parameters of the system, namely, the Larmor radius, the Larmor frequency and the amplitude of the electrostatic potential. We assume that the characteristic time scale and spatial scale of the turbulent potential are rescaled to $2\pi$ without loss of generality and for simplicity (see Supplemental Material at [URL] for details on the nondimensionalization). The main parameters of the system are
\begin{eqnarray*}
&& A=\Phi_0/B,\\
&& \rho = \frac{ \sqrt{2 k_B T /m}}{\vert \Omega \vert},\\
&& \eta = \frac{1}{2 \Omega},
\end{eqnarray*}
where $T$ is the temperature in the direction perpendicular to magnetic field lines and $\Phi_0$ is the amplitude of the electrostatic potential $\Phi(x,y,t)$. Effectively, $A$ is the amplitude of a potential $\phi(x,y,t)$ = $\Phi(x,y,t)/B$, which is the one governing the dynamics of charged particles. In the plane perpendicular to the magnetic field lines, the rescaled equations of motion become $\dot{\bf x}=\rho {\bf v}/(2\vert \eta\vert)$ and $\dot{\bf v}=-{\rm sgn}(\eta) \nabla \phi /\rho +{\bf v}\times \hat{\bf z} / (2\eta)$. The resulting Hamiltonian system has two and a half degrees of freedom (one degree of freedom in each direction perpendicular to the magnetic field and half a degree of freedom for the explicit time dependence of the electrostatic potential). In addition, the typical (fast) time scale of the dynamics is $\pi \vert \eta\vert$. The phase space of the particle is of dimension 5, which does not allow for a facilitated visualization of the phase space structures responsible for transport properties. 

In order to reduce the dynamics (see Supplemental Material at [URL] for a brief description of the assumptions and ordering used in the reduction), we decouple the fast from the slow temporal scales, by using the guiding-center theory in a Hamiltonian setting~\cite{Littlejohn1979,Littlejohn1983,Cary2009,Brizard2007}. The main ingredient is a change of positions from the particles (at position ${\bf x}$) to the guiding centers (at position ${\bf X}$) defined by ${\bf x}={\bf X}+\hat{\bf z}\times{\bf v}/\Omega$ at the lowest order. The fast oscillations are generated by the term $\hat{\bf v}\times{\bf z}$ in the equation for $\dot{\bf v}$ which can be seen by introducing a gyroangle $\theta$ which rotates with a frequency $\Omega$. Using the guiding-center positions as variables and performing a suitable change of coordinates (using, e.g., Lie transforms) to eliminate the $\theta$-dependence in the Hamiltonian at the lowest orders, the dynamics is reduced to the motion of effective particles (guiding centers) subjected to an ${\bf E}\times {\bf B}$ drift velocity in an effective electrostatic potential $\psi(X,Y,t)$ which depends parametrically on $A$, $\rho$ and $\eta$, and is given by~\cite{Cary2009,Brizard2007}
$$
\psi = \mathbb{J}_0[\phi]- \eta\left(\mathbb{J}_1[\phi^2] -2 \mathbb{J}_0[\phi]\mathbb{J}_1[\phi]\right),
$$
where $\mathbb{J}_0$ is the gyro-average operator defined by
$$ 
\mathbb{J}_0[\phi](X, Y, t;\rho)=\frac{1}{2\pi}\int_0^{2\pi}  \phi\left( X+\rho \cos\theta, Y-\rho \sin\theta , t\right){\rm d}\theta,
$$
and $\mathbb{J}_1[\phi]=\rho^{-1}(\partial/\partial\rho)\mathbb{J}_0[\phi]$.
The dynamics of the guiding centers is driven by the ${\bf E}\times {\bf B}$ drift where an effective electric field is generated by the effective potential $\psi(X,Y,t)$, i.e., $\dot{\bf X}= -\nabla \psi \times \hat{\bf z}$ in the rescaled units. 
We notice that we have used the guiding-center reduction at the second order in the amplitude of the electrostatic potential in order to have all three parameters of the particle dynamics, namely $A$, $\rho$ and $\eta$, present in the reduced equations for the guiding-center dynamics. At first order where the effective potential $\psi$ is given by $\mathbb{J}_0[\phi]$, the equations of motion are independent of $\eta$, preventing the study of the influence of this parameter in the dynamics. 

The main advantage of using the guiding-center dynamics is that the fast dynamics of the velocities of the particles are decoupled from the slow motion of the guiding centers, and allows for the use of larger time steps which greatly facilitates numerical simulations. In addition, this reduces the dimensionality of the Hamiltonian system to one and a half degree of freedom, namely, the $(X,Y)$ degree of freedom in addition to the explicit time dependence. The phase space of the guiding centers is of dimension 3 which allows for a facilitated visualization of phase space structures using, e.g., Poincar\'e sections. Here we take advantage of this reduction to identify the phase-space structures governing the transport properties in the system.   

In order to model the turbulent electrostatic potential, we choose the following electrostatic potential in the rescaled units~\cite{Pettini1988}:
$$
\phi(x,y,t)= A \sum_{\substack{n,m=1 \\ n^2+m^2\leq M^2}}^M \frac{1}{(n^2+m^2)^{3/2}} \sin(n x +m y + \varphi_{nm} -t),
$$
where $\varphi_{nm}$ are random phases (uniform distribution in $[0,2\pi[$). Together with the decrease of the amplitude associated with small scales of typical size $2\pi/k$ as $k^{-3}$, this potential mimics some of the features of a turbulent electrostatic potential, notably electrostatic drift-wave turbulence.

\begin{figure}
    \centering
    \includegraphics[width=13cm]{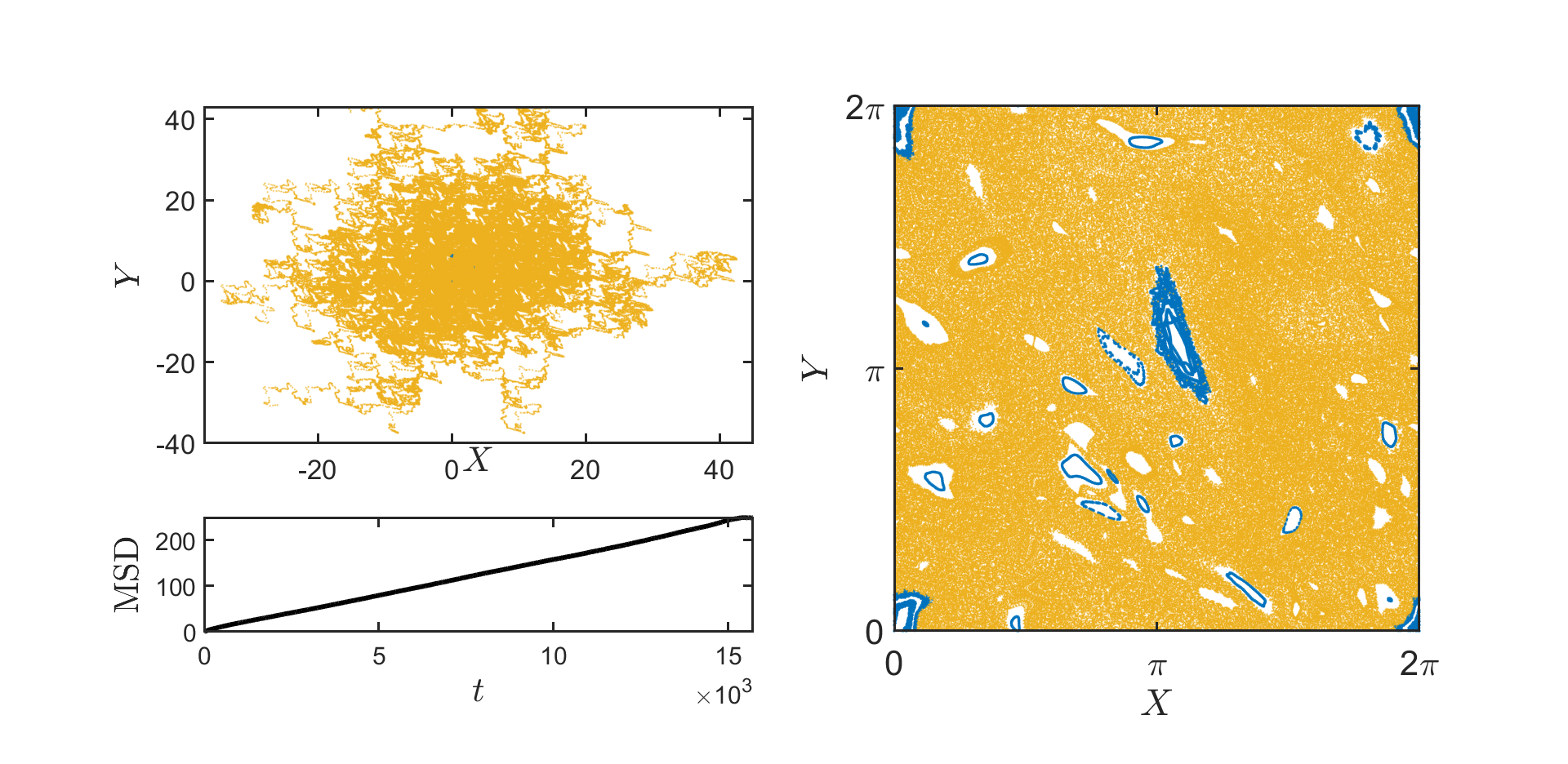}
    \caption{Upper left panel: Poincar\'e section in $\mathbb{R}^2$ of guiding-center trajectories. Lower left panel: Values of MSD of guiding centers as a function of time. Right panel: Poincar\'e section in $(\mathbb{R}/(2\pi\mathbb{Z}))^2$ of guiding-center trajectories. The blue (black) dots correspond to trapped particles. The light orange (light gray) dots correspond to chaotic trajectories.  The parameters are $A = 0.7$, $\eta = 0$ and $\rho = 0$. All units are dimensionless.}\label{fig:fig1}
\end{figure}

In what follows, we fix $M=25$ and $A=0.7$ and vary the other two parameters $\rho$ and $\eta$. For each values of the parameters, we integrate numerically the equations of motion for the guiding centers for a large ensemble of initial conditions in $[0, 2\pi[^2 $ (see Supplemental Material at [URL] for a brief description of the numerical scheme~\footnote{The numerical code (in Python) to integrate the particle dynamics and the guiding-center dynamics is available at \url{https://github.com/cchandre/Guiding-Center}.}). As it has already been described in the literature, for $\rho=\eta=0$, the dynamics exhibit two main types of trajectories: the trapped ones which remain inside elliptic islands forever, and chaotic ones which resembles stochastic diffusion. The latter ones contribute the most to transport properties, of diffusive type. In Fig.~\ref{fig:fig1}, we represent the expected diffusive (chaotic) dynamics and the trapped particles (upper left panel). The diffusive character is evidenced by computing the time- and ensemble-averaged mean square displacement (MSD) $\langle r^2(t)\rangle$ of a set of untrapped trajectories (lower left panel) (see Supplemental Material at [URL] for its explicit expression). In order to better visualize the phase-space structures, we plot a Poincar\'e section (stroboscopic plot), i.e., the positions $(X(2n\pi),Y(2n\pi))$ for $n\in \mathbb{N}$, of the guiding centers at each period of the field (right panel). The Poincar\'e section clearly evidences the chaotic dynamics of diffusive particles and the regular motion associated with the trapped particles. 

\begin{figure}
    \centering
    \includegraphics[width=13cm]{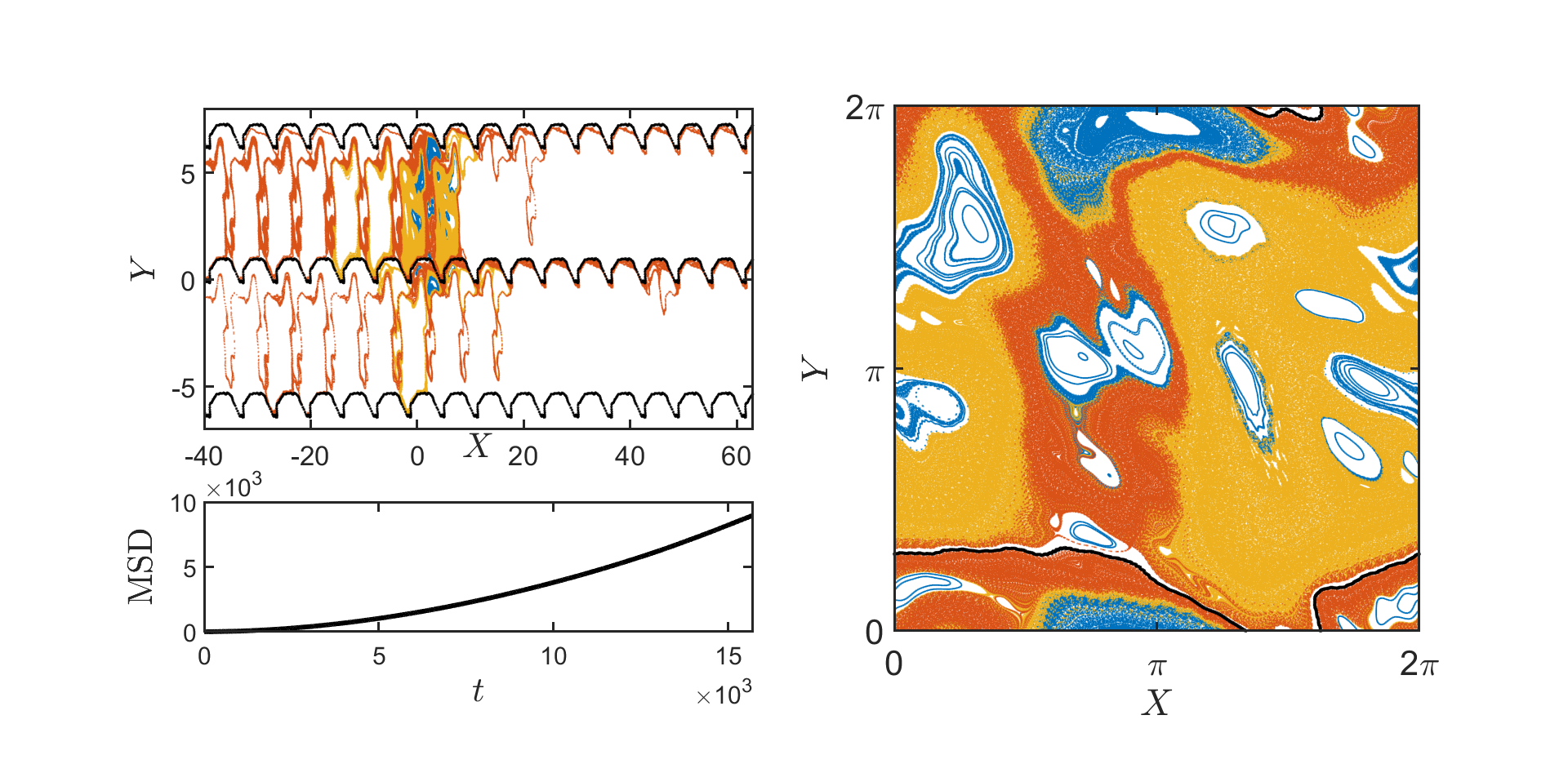}
    \caption{Upper left panel: Poincar\'e section in $\mathbb{R}^2$ of guiding centers trajectories. Lower left panel: Values of MSD of guiding centers as a function of time. Right panel: Poincar\'e section in $(\mathbb{R}/(2\pi\mathbb{Z}))^2$ of guiding-center trajectories. The continuous black line on the right panel and on the upper left panel corresponds to Poincar\'e sections of the twistless invariant torus organizing the lower layer of super-diffusive transport. The blue (black) dots correspond to trapped particles. The light orange (light gray) dots correspond to chaotic trajectories. The dark orange (dark gray) dots correspond to ballistic trajectories. The parameters are $A = 0.7$, $\eta = 0.14$ and $\rho = 0.3$.}\label{fig:fig2}
\end{figure}

We now increase $\rho$ to investigate its role in the dynamics. For potentials with few spatial Fourier modes, it was shown~\cite{Manfredi1996,Manfredi1997,daFonseca2016,Kryukov2018} that the main effect is to reduce diffusion. In particular, it was shown in Ref.~\cite{Kryukov2018} that the effect of increasing the Larmor radius was to regularize the dynamics by decreasing the effective amplitude of the electrostatic potential (i.e., $A$ was replaced by $A J_0(\rho\sqrt{2})$ with the Bessel function of the first kind $J_0$). Here we show that the role of $\rho$ is more subtle when a spatial structure of the electrostatic potential is introduced.   
In Fig.~\ref{fig:fig2}, we represent the dynamics of guiding centers in $\mathbb{R}^2$ (upper left panel) and the Poincar\'e section in $(\mathbb{R}/(2\pi\mathbb{Z}))^2$ (right panel) for $A=0.7$, $\rho=0.3$ and $\eta=0.14$. We notice the same two types of trajectories as in Fig.~\ref{fig:fig1}, namely the trapped and the chaotic trajectories. The main difference with Fig.~\ref{fig:fig1} is that a new type of trajectories emerges, very elongated in one direction. The MSD $\langle r^2(t)\rangle$ shown in the lower left panel of Fig.~\ref{fig:fig2} displays a close-to-quadratic behavior in time, indicating a super-diffusive/ballistic behavior.  

\begin{figure}
    \centering
    \includegraphics[width=13cm]{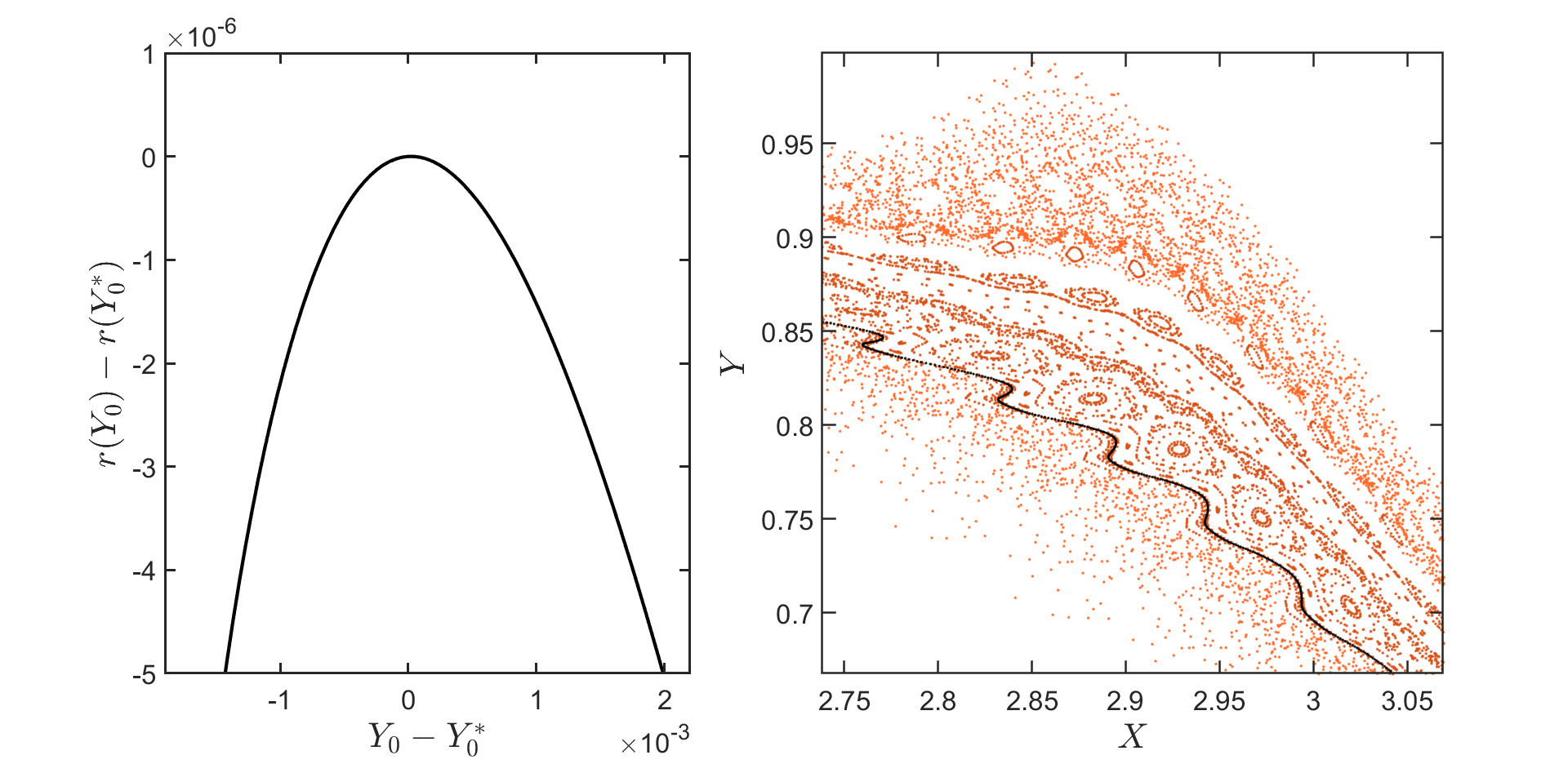}
    \caption{Left panel: Weighted-Birkhoff averages of rotation numbers computed for regular structures as a function of the initial condition $Y_0$ while $X_0 = \pi$, the reference values are $Y_0^* = 0.58$ and $r ( Y_0^* ) \approx 0.087647$. The continuous black line on the left panel is an inset of the twistless invariant torus depicted in Fig.~\ref{fig:fig2}. Right panel: Inset of Poincar\'e section shown in Fig.~\ref{fig:fig2}. The parameters are $A = 0.7$, $\eta = 0.14$ and $\rho = 0.3$.}\label{fig:fig3}
\end{figure}

\begin{figure}
    \centering
    \includegraphics[width=13cm]{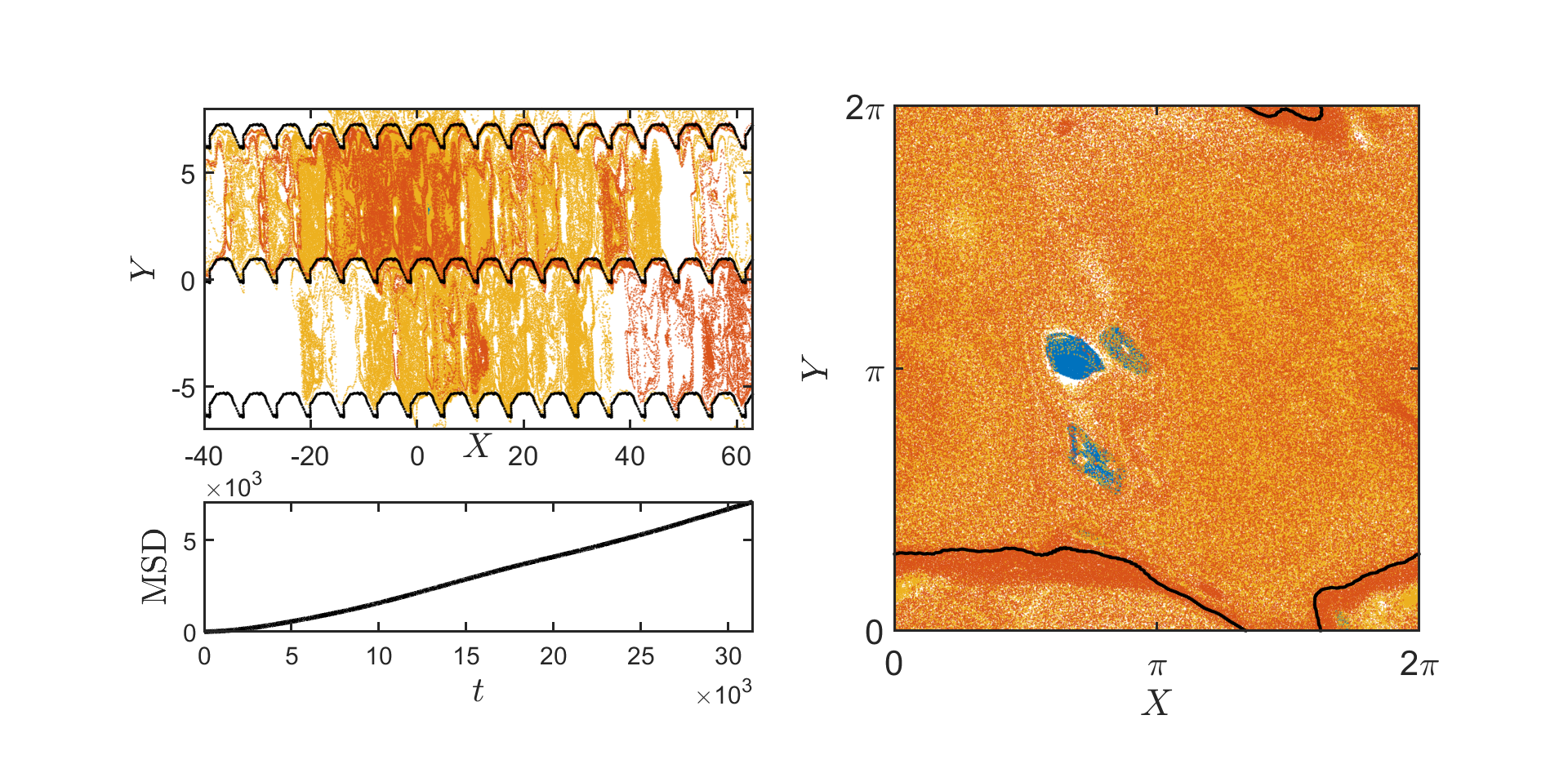}
    \caption{Upper left panel: Poincar\'e section in $\mathbb{R}^2$ of guiding centers trajectories computed from full-orbit trajectories. Lower left panel: Values of MSD of guiding centers as a function of time. Right panel: Poincar\'e section in $(\mathbb{R}/(2\pi\mathbb{Z}))^2$ of guiding-center trajectories computed from full-orbit trajectories. The blue (black) dots correspond to trapped particles. The light orange (light gray) dots correspond to chaotic trajectories. The dark orange (dark gray) dots correspond to ballistic trajectories. The black lines in the upper left panel and in the right panel indicate the twistless invariant torus found in the guiding-center approximation (same as in Fig.~\ref{fig:fig2}). The parameters are $A = 0.7$, $\eta = 0.14$ and $\rho = 0.3$.}\label{fig:fig4}
\end{figure}

\begin{figure}
    \centering
    \includegraphics[width=12cm]{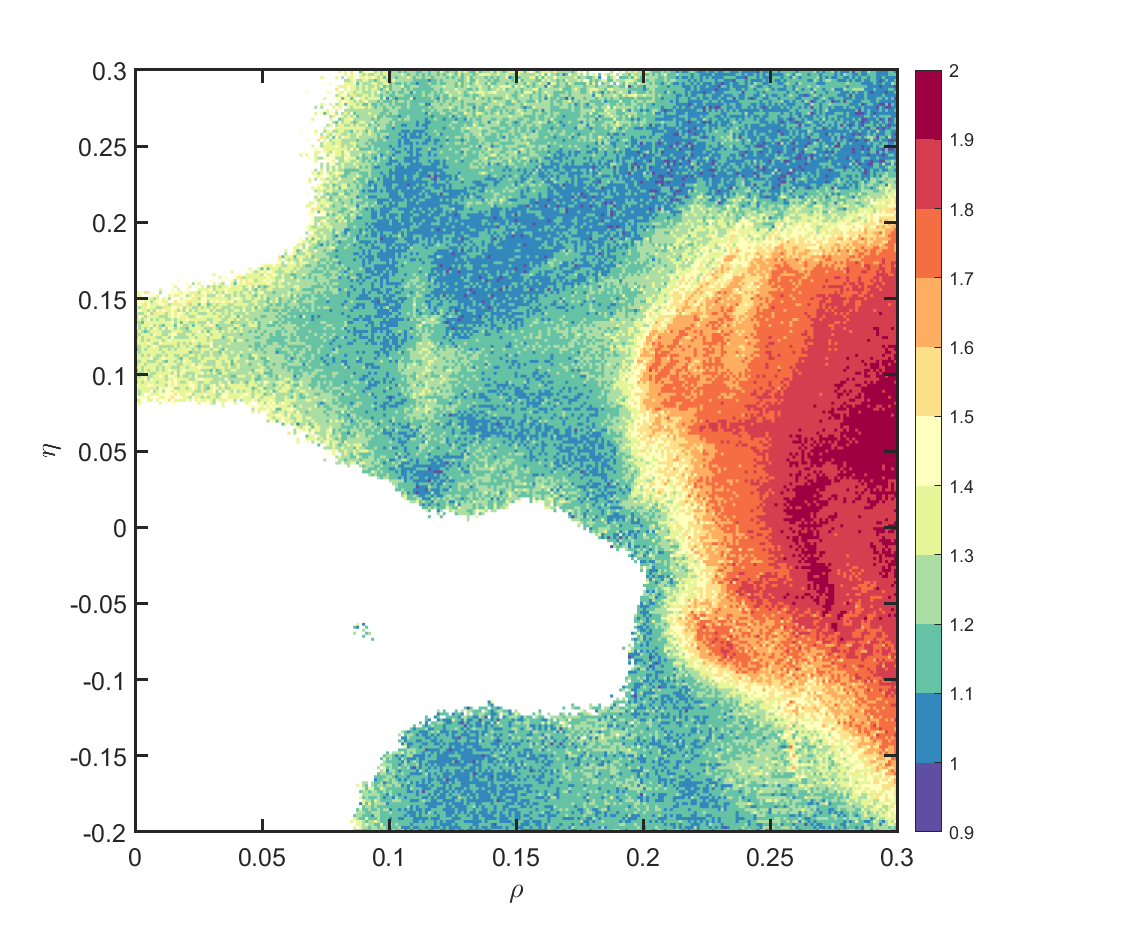}
    \caption{Values of $b$ extracted from a power law interpolation $ (a t)^b $ of the values of MSDs of guiding centers (obtained from guiding-center dynamics) as functions of time $t$ for different values of the parameters $\rho$ and $\eta$. The white region is where no significant super-diffusive behavior was observed. $A = 0.7$ is fixed for all the cases. }\label{fig:fig5}
\end{figure}

By looking at the Poincar\'e section in the right panel of Fig.~\ref{fig:fig2} we notice that the trajectories leading to this super-diffusive behavior are all organized in rather thin layers. A zoom of one of these layers is displayed in Fig.~\ref{fig:fig3}. We clearly see that this region is organized in invariant tori and resonant islands of rather large periods, evidencing some regular structures as responsible for the super-diffusive behavior. In order to get more insights into this region, we compute the rotation numbers of these regular structures. In the left panel of Fig.~\ref{fig:fig3}, we display the weighted-Birkhoff averages for these rotation numbers~\cite{Sander2020} as a function of the initial condition $Y_0$ (see Supplemental Material at [URL] for their explicit expression). These rotation numbers are on a bell-shaped curve, clearly evidencing the presence of a twistless invariant torus~\cite{delCastillo1996,Fuchss2006,Martinell2013} at the center of the region where super-diffusive behavior occurs.  
It should be noticed that these invariant structures constitute barriers of transport in the $Y$-direction while drastically enhancing transport in the $X$-direction. More precisely, Fig.~\ref{fig:fig2} displays two regions of super-diffusive transport, one containing a twistless invariant torus, while another containing the remnants of a broken one. The first leads to a super-diffusion in the positive $X$ direction, the second one in the negative $X$ direction, as it can be seen in the upper left panel in Fig.~\ref{fig:fig2}. It has also been verified that for $\rho \gtrsim 0.5$, there are two invariant twistless tori, one with a positive rotation number and one with a negative one (see Supplemental Material at [URL] for a Poincar\'e section), meaning that the broken twistless invariant tori has been restored by increasing $\rho$.  
Moreover, given the shape of the upper super-diffusive layer, the region of diffusive transport is pinched, and therefore the diffusive behavior is almost completely suppressed, only a few particles diffuse through the holes of the broken invariant structure, so extremely slowly. Transport properties are dominated by this super-diffusive behavior. 

In Fig.~\ref{fig:fig4}, we display a Poincar\'e section of the guiding centers reconstructed from the full orbits obtained with Eq.~\eqref{eqn:fo} (see Supplemental Material at [URL] for the reconstruction method). We notice that some of the structures present in the guiding-center dynamics can still be observed and in particular, the regions where super-diffusive behavior occurs. This observation validates the conclusions drawn using the guiding-center approximation.

The importance of twistless invariant tori resides in their robustness with respect to perturbation, much more robust than regular invariant tori, as present, e.g., in trapped islands (blue regions). As a consequence of their robustness, we expect their presence in a rather large region in parameter space. 

For each values of the parameters $(\rho,\eta)$, we compute the values of MSD of guiding centers as a function of time and interpolate these values with a power law, i.e., ${\rm MSD}(t)\approx (at)^b$.  
In Fig.~\ref{fig:fig5}, we represent the map of the values of $b$ in parameter space $(\rho,\eta)$. We notice that a large region of super-diffusive behavior is present for $\rho \gtrsim 0.2$ and $\vert \eta\vert \lesssim 0.2$. Poincar\'e sections confirm that this large region of super-diffusive behavior is due to the presence of two twistless invariant tori or remnants of broken twistless invariant tori.  
The transition toward a super-diffusive/ballistic behavior occurs at around $\rho\approx$ 0.2--0.25 which corresponds to 3 to 4\% of the typical length scale of the electrostatic potential.  

As an example, Refs.~\cite{Sabot2006, Hennequin2006, Casati2009} provide typical values of $e\times \Phi/T=1\%$, $\Omega\tau=9.3\times 10^3$ and $\lambda/\rho=20$ in the Tore Supra tokamak. These values correspond to dimensionless parameters $\rho=0.3$, $\eta\simeq 7\times 10^{-5}$, $A=0.7$. 
As shown in Fig.~\ref{fig:fig5}, these values are well inside the red region where a super-diffusive behavior is expected (see Supplemental Material at [URL] for a Poincar\'e section). 

The presence of twistless invariant tori in tokamak plasma physics has been previously advocated for magnetic configurations presenting locally a reversed shear in their safety factor profile~\cite{Balescu1998}. Here the source of creation of such twistless invariant tori is completely different since there is no shear in the magnetic configuration. The origin of the resulting transport barrier is solely a consequence of the electrostatic turbulence, and more precisely of the spatial structure of the electrostatic potential.

Anomalous transport was observed in electrostatic drift-wave turbulence (see, e.g., Refs.~\cite{Balescu1998,Annibaldi2000,Gustafson2008}) by tweaking the electromagnetic configuration or the equilibrium density of the particles. Here the main result is that, with the same electric and magnetic field, the nature of transport of charged particles can be completely different for different particles. We identified a transition from diffusive to super-diffusive behavior in the plane perpendicular to the magnetic field as the Larmor radius is increased. This super-diffusive behavior is due to the presence of twistless invariant tori which constitute robust barriers of transport in one spatial direction and is associated with ballistic transport in the other spatial direction.   

\begin{acknowledgments}
M.S.\ and F.A.\ contributed equally to this work. Centre de Calcul Intensif d’Aix-Marseille is acknowledged for granting access to its high performance computing resources. 
This work has been carried out within the framework of the French Federation for Magnetic Fusion Studies (FR-FCM).
\end{acknowledgments}

\end{document}